\documentclass[12pt]{article}

\usepackage[a4paper,left=30mm,right=20mm,top=20mm,bottom=20mm]{geometry}
\usepackage{graphics}
\usepackage{amsmath}
\usepackage{epsfig}
\usepackage{cite}
\usepackage{xcolor}

\newcommand{\z}{&&\hspace*{-1cm}}

\newcommand{\bea}{\begin{eqnarray}}
\newcommand{\eea}{\end{eqnarray}}
\newcommand{\be}{\begin{equation}}
\newcommand{\ee}{\end{equation}}

\title{Pion-photon transition form factor with analytic coupling}

\author{I.A.~Zemlyakov$^{1,2}$, I.L. Chuev$^{3}$, A.V.~Kotikov$^{4}$}
\date{\vspace{-5ex}}
\begin{document}

\maketitle
\begin{center}
  {\it $^1$Department of Physics, Universidad Tecnica Federico Santa Maria,\\
  Avenida Espana 1680, Valparaiso, Chile,}\\
  {\it $^2$Tomsk State University, 634010 Tomsk, Russia}\\
  {\it $^3$Department of Theoretical Physics,
  Moscow Institute for Physics and Technology,
  141701 Dolgoprudny, Russia}\\
  {\it $^4$Bogoliubov Laboratory of Theoretical Physics,
  Joint Institute for Nuclear Research, 141980 Dubna, Russia}\\

\end{center}


\begin{center}

{\bf Abstract }

\end{center}

  We investigate the pion-photon transition form factor
  within the framework of analytic QCD.
  A comparison is performed between experimental data
  and perturbative QCD
  based on conventional and analytic versions of perturbation theory with the
``massive'' form of the
twist-four contribution. We show that conventional perturbation
theory 
fails to reproduce the data, while
the analytic version
demonstrates good agreement with experiment.

\section{Introduction}

The pion-photon transition form factor (TFF) is
used to study the chiral symmetry, the quark-mass ratio, the characteristics of the pseudo-scalar meson's decay, and etc..
At the twist-two level,
the pion-photon TFF with one real and one virtual photon
can be decomposed into two parts: the perturbatively calculable coefficient function (CF) and the non-perturbative
twist-two pion light-cone distribution amplitude (LCDA)~
\cite{Efremov:1979qk,Lepage:1979zb}

The valence
quark part of TFF
provides the dominant contribution and here
we limit ourselves to considering only this term. It
twist-two part 
$Q^{2} F_{\rm V}^{\pi \gamma}(Q^{2})$ can be expressed as follow~\cite{Efremov:1979qk,Lepage:1979zb}
\begin{eqnarray}
Q^{2} F^{\pi \gamma (\tau=2)}_{\rm V}(Q^{2}) &=& \frac{\sqrt{2}f_{\pi}}{6}\int^{1}_{0}dx\, T(x,Q,\mu_{f})\, \phi_{\pi}(x,\mu_{f}),
\label{eq1}
\end{eqnarray}
where the pion decay constant $f_{\pi}=130.5$ MeV~\cite{ParticleDataGroup:2022pth} and the pion LCDA
$\phi_{\pi}(x,\mu_{f})$
is usually represented as a Gegenbauer polynomial expansion~\cite{Efremov:1979qk,Lepage:1979zb},
e.g.,
\begin{eqnarray}
\phi_{\pi}(x,\mu_{f})&=&6x\overline{x}\sum_{n=0,2,\cdots}b_{n}(\mu_f) C^{3/2}_{n}(2x-1),
\label{eq2}
\end{eqnarray}
where $\overline{x}=1-x$, $C^{3/2}_{n}(2x-1)$ are Gegenbauer polynomials and $\mu_{f}$ is the factorization scale. The first moment $b_{0}$ equals to $1$ by
using the normalization condition.

Many efforts have been devoted to the CF calculations.
The complete next-to-next-to-leading-order (NNLO) QCD corrections using the conformal symmetry~\cite{Braun:2021grd} and the hard-collinear factorization
theorem~\cite{Gao:2021iqq} have been reported rather recently. 
At NNLO accuracy the
CF can be written at NNLO accuracy as
\begin{eqnarray}
  T(x,Q) &=& T^{(0)}(x)+a_s(Q) T^{(1)}(x)
  +a_s^{2}(Q) T^{(2)}(x) + {\cal O}(a_s^{3}),
\label{eq3}
\end{eqnarray}
where hereafter we use $\mu_{r}=\mu_{f}=Q$ for the renormalization and factorizaion scales and $a_s=\beta_0\alpha_{s}/(4\pi)$.

We use several 
sets of the Gegenbauer moments at  $Q_{0}=1$ GeV:
\bea
&& b_{2}(Q_{0})=0.159,~~b_{4}(Q_{0})=-0.098,~~\mbox{
  MPS set \cite{Mikhailov:2021znq}};\label{LA}\\
&& b_{2}(Q_{0})=0.203,~~b_{4}(Q_{0})=-0.143,~~\mbox{BMS set \cite{Bakulev:2001pa}};\label{BMS}\\
&& b_{2}(Q_{0})=0.112,~~b_{4}(Q_{0})=-0.029,~~\mbox{MPS$_2$ set \cite{Mikhailov:2021znq}};\label{MPS}\\
&& b_{2}(Q_{0})=0.0812,~~b_{4}(Q_{0})=-0.0191,~~\mbox{SP set \cite{Stefanis:2014nla}};\label{SP}
\eea
A fairly long list of values of $b_{2}$ and $b_{4}$ can be found, for example, in \cite{Bakulev:2003cs}.


\section{Transition form factor}
\label{sec:mesonic-FAPT}

Keeping only first three terms in (\ref{eq2}) and
evaluating the integrals in r.h.s. of (\ref{eq1}),  
the perturbative expansion of the valence twist-two TTF part 
$Q^2 F^{\gamma \pi (\tau=2)}_\text{V}$ can be expressed as
\be
F^{\gamma \pi(\tau=2)}_\text{V}\left(Q^2\right)
=  F^{\gamma \pi(\tau=2)}_\text{V,n=0}\left(Q^2\right)+\hat{b}_2(Q_0^2)\, F^{\gamma \pi(\tau=2)}_\text{V,n=2}\left(Q^2\right) + \hat{b}_4(Q_0^2)\,
F^{\gamma \pi(\tau=2)}_\text{V,n=4}\left(Q^2\right)\,,
\label{TFF1a}
\ee
where
\be
\hat{b}_n(Q_0^2)=\frac{b_n(Q_0^2)}{a^{d_n}_s(Q_0^2)}
\label{ha_n}
\ee
and $b_n(Q_0^2)$ are given in (\ref{LA})-(\ref{SP}). 

The perturbative part of $Q^2 F^{\gamma \pi(\tau=2)}_\text{V}\left(Q^2\right)$ can be expressed as
\bea
&&Q^2 F^{\gamma \pi}_\text{V,n=0}\left(Q^2\right)=r_0^{[0]}+r_1^{[0]}a_s(\mu^2)+(r_2^{[0]}+\beta_0\overline{r}_2^{[0]})a^2_s(\mu^2)+O(a^3_s)\,,\label{TFF_2}\\
&&Q^2 F^{\gamma \pi(\tau=2)}_\text{V,n}\left(Q^2\right)= a^{d_n+1}_s(Q^2)\left(r_1^{[n]}+(R_2^{[n]}+\beta_0\overline{R}_2^{[n]})a_s(Q^2)+O(a^2_s)\right)\,,\label{TFF_n_2a}
\eea
where
\be
d_{n=0}=0,~~d_{n=2}=\frac{50}{81},~~d_{n=4}=\frac{364}{405}\,,
\label{gamma0a}
\ee
for $f=3$ and, thus, $\beta_0=9$.

The coefficients in (\ref{TFF_2}) and (\ref{TFF_n_2a}) are (see \cite{Zhou:2023ivj}):
\bea
\z r_0^{[0]}=0.185,~~\beta_0 r_1^{[0]}=-1.230,~~~\beta^2_0 r_2^{[0]}=-7.015,~~\beta^2_0 \overline{r}_2^{[0]}=-2.674,~~\beta_0 r_1^{[2]}=0.208
\,,\nonumber \\
\z
\beta^2_0 R_2^{[2]}=-7.992,~~\beta^2_0 \overline{R}_2^{[2]}=0.995,~~
\beta_0 r_1^{[4]}=0.135,~~\beta^2_0 R_2^{[4]}=-1.699,~~\beta^2_0 \overline{r}_2^{[2]}=0.510
\,.~~~~~~
\label{ri}
\eea

When we moved from powers $a^{d}_s(Q^2)$ to derivatives $\tilde{a}_{d}(Q^2)$ (see \cite{Kotikov:2022JETP} and discussions therein),
the coefficients in
(\ref{TFF_2}) and (\ref{TFF_n_2a}) are
not changed (at the considered level of accuracy), i.e.
\bea
&&Q^2 F^{\gamma \pi(\tau=2)}_\text{V,n=0}\left(Q^2\right)=r_0^{[0]}+r_1^{[0]}a_s(Q^2)+(r_2^{[0]}+\beta_0\overline{r}_2^{[0]})\tilde{a}_2(Q^2)+O(\tilde{a}_{3})\,,
\label{TFF_0_NNLOb}\\
&&Q^2 F^{\gamma \pi(\tau=2)}_\text{V,n}\left(Q^2\right)= r_1^{[n]}\tilde{a}_{d_n+1}(Q^2)+(R_2^{[n]}+\beta_0\overline{R}_2^{[n]})\tilde{a}_{d_n+2}(Q^2)+O(\tilde{a}_{d_n+3}) \,.\label{TFF_n_2b}
\eea

In the framework of the analytic perturbation theory (APT),
we have
\be
F^{\gamma \pi(\tau=2)}_\text{V,A}\left(Q^2\right)
= F^{\gamma \pi(\tau=2)}_\text{V,A,n=0}\left(Q^2\right)+\hat{b}_2(Q_0^2)\, F^{\gamma \pi (\tau=2)}_\text{V,A,n=2}\left(Q^2\right)
+\hat{b}_4(Q_0^2)\, F^{\gamma \pi(\tau=2)}_\text{V,A,n=4}\left(Q^2\right)\,,
\label{TFF1A}
\ee
where $\hat{b}_2(Q_0^2)$ and $\hat{b}_4(Q_0^2)$ are given in (\ref{ha_n}) and
\bea
&&Q^2 F^{\gamma \pi(\tau=2)}_\text{V,A,n=0}\left(Q^2\right)=r_0^{[0]}+r_1^{[0]}A_1(Q^2)+(r_2^{[0]}+\beta_0\overline{r}_2^{[0]})\tilde{A}_2(Q^2)+O(\tilde{A}_{3})\,,\label{TFF_0_NNLO_A}\\
&&Q^2 F^{\gamma \pi(\tau=2)}_\text{V,A,n}\left(Q^2\right)= r_1^{[n]}\tilde{A}_{d_n+1}(Q^2)+(R_2^{[n]}+\beta_0\overline{R}_2^{[n]})\tilde{A}_{d_n+2}(Q^2)+O(\tilde{A}_{d_n+3}) \,.\label{TFF_n_2b_A}
\eea
The results for the analytic couplings $\tilde{A}_d(Q^2)$ can be found in Ref. \cite{Kotikov:2022sos}. 

When we added higher-twist corrections (in the massive form, according to \cite{Teryaev:2013qba}) to the parts of
$F^{\gamma \pi(\tau=2)}_\text{V,n=0}\left(Q^2\right)$ and $F^{\gamma \pi(\tau=2)}_\text{V,A,n=0}\left(Q^2\right)$
(as was done in \cite{Gabdrakhmanov:2023rjt} and \cite{Gabdrakhmanov:2025vxw} for the polarized Bjorken sum rule (BSR) and the Gross-Llewellyn-Smith (GLS) sum rule, respectively),
they take the form
\bea
&&Q^2 F^{\gamma \pi}_\text{V,n=0}\left(Q^2\right)=  F^{\gamma \pi(\tau=2)}_\text{V,n=0}\left(Q^2\right)
+\frac{\mu_4(Q^2) \tilde{M}^2}{Q^2+\tilde{M}^2}\,,
\label{TFF_0_NNLOc}\\
&&Q^2 F^{\gamma \pi}_\text{V,A,n=0}\left(Q^2\right)= F^{\gamma \pi(\tau=2)}_\text{V,A,n=0}\left(Q^2\right)
+\frac{\mu_{\rm A,4}(Q^2) M^2}{Q^2+M^2}
\,.\label{TFF_0_NNLO_A_HT}
\eea

For $\mu_4(Q^2)$ and $\mu_{\rm A,4}(Q^2)$ we use two possibilities: $Q^2$-independent values and  $Q^2$-dependent ones with
\be
\mu_4(Q^2)=\mu_4(Q_0^2)\frac{a_s^{\nu}(Q^2)}{a_s^{\nu}(Q_0^2)}\,,~~
\mu_{\rm A,4}(Q^2)=\mu_{\rm A,4}(Q_0^2)\frac{A_{\nu}(Q^2)}{a_s^{\nu}(Q_0^2)},~~\nu=\frac{\gamma^{(\tau=4)}}{\beta_0}=\frac{32}{81} \,,
\label{Tw4A}
\ee
where $\gamma^{\tau=4}=32/9$ is the twist-four anomalous dimension and $Q_0$=1 GeV.

\begin{figure}[t]
\centering
\includegraphics[width=0.68\textwidth]{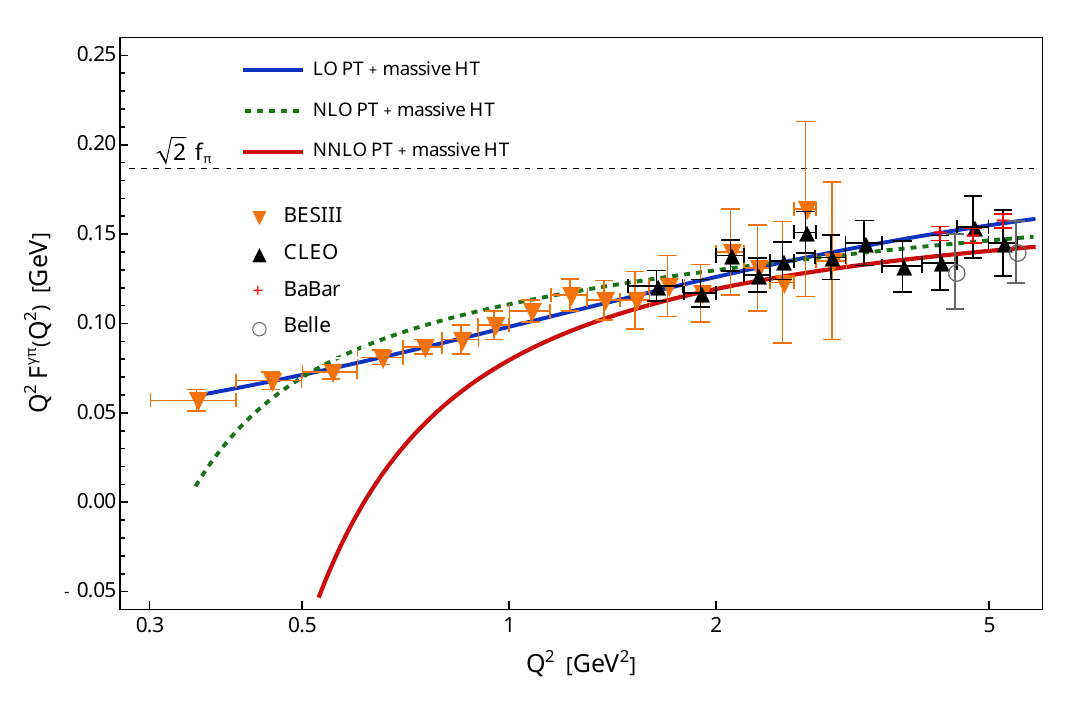}
\caption{\label{fig:as1352}
  The results (\ref{TFF1a}), (\ref{TFF_0_NNLOb}), (\ref{TFF_n_2b}) and (\ref{TFF_0_NNLOc}) in the first three orders of ordinary QCD with $b_n(Q_0^2)$ given in (\ref{LA}).
}
\end{figure}

\begin{figure}[!htb]
\centering
\includegraphics[width=0.68\textwidth]{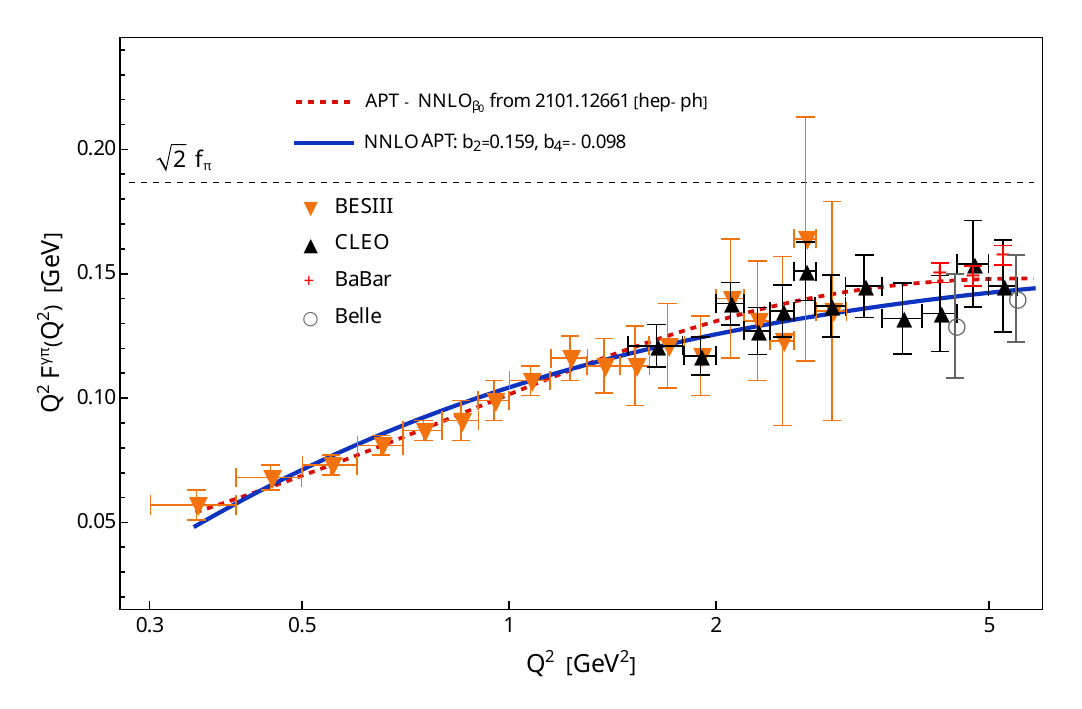}
\caption{\label{fig:as1352}
  The results of (\ref{TFF1A}), (\ref{TFF_0_NNLO_A}), (\ref{TFF_n_2b_A}) and (\ref{TFF_0_NNLO_A_HT}), where $b_n(Q_0^2)$ are given in (\ref{LA}), in
  NNLO APT together with the results obtained in \cite{Mikhailov:2021znq}.
}
\end{figure}

\section{Results}

By analogy with the results in Refs. \cite{Gabdrakhmanov:2023rjt,Gabdrakhmanov:2025vxw,Ayala:2017uzx}, where the BSR and GLS cases were considered,
we fit the experimental data for the pion-photon TFF within the framework of standard and analytic QCD with the
``massive'' form of the twist-four term (see Eqs. (\ref{TFF1a}), (\ref{TFF_0_NNLOb})-(\ref{TFF_0_NNLO_A_HT}) above).

For conventional QCD, Fig. 1 shows a discrepancy between the QCD predictions and the experimental data (references to the papers with experimental data can be found
in \cite{Mikhailov:2021znq}). Indeed, the leading-order (LO) results agree well with the experimental data, but beyond LO, the discrepancy increases with
increasing perturbation theory (PT) order.
This is the same as in the BSR and GLS cases (see \cite{Gabdrakhmanov:2023rjt,Gabdrakhmanov:2025vxw}), and the reason is the same.
With increasing PT order, the Landau pole of the strong coupling shifts toward higher values of $Q^2$
(see, for example, Fig. 1 in \cite{Kotikov:2022sos}) and hinders the PT applicability.

\begin{table}[t]
\begin{center}
\begin{tabular}{|c|c|c|c|}
\hline
& $M^2$ [GeV$^2$] & $\mu_{\rm{A},4}$  & $\chi^2/({\rm d.o.f.})$   \\
\hline 
LO & 1.116 $\pm$ 0.123 (1.116 $\pm$ 0.123) & -0.165 $\pm$ 0.008 (-0.165 $\pm$ 0.008) & 0.42 (0.42)  \\
& \{1.116 $\pm$ 0.123\} [1.116 $\pm$ 0.123] & \{-0.165 $\pm$ 0.008\} [-0.165 $\pm$ 0.008] & \{0.42\} [0.42]  \\
\hline
  NLO & 0.381 $\pm$ 0.091 (0.382 $\pm$ 0.091) & -0.183 $\pm$ 0.027 (-0.183 $\pm$ 0.027) & 0.55 (0.55) \\
 & \{0.380 $\pm$ 0.091\} [0.379 $\pm$ 0.091] & \{-0.183 $\pm$ 0.027\} [-0.183 $\pm$ 0.028] & \{0.55\} [0.55] \\
  \hline
  N$^2$LO & 0.119 $\pm$ 0.074 (0.119 $\pm$ 0.074) & -0.311 $\pm$ 0.161 (-0.309 $\pm$ 0.159)& 1.20 (1.20)  \\
  & \{0.118 $\pm$ 0.074\} [0.117 $\pm$ 0.074] & \{-0.311 $\pm$ 0.161\} [-0.313 $\pm$ 0.165])& \{1.20\} [1.21]  \\
 \hline
\end{tabular}
\end{center}
\vspace{-0.3cm}
\caption{%
  The fit parameters for the $Q^2$-independent $\mu_{\rm{A},4}$ with $b_n(Q_0^2)$ given in (\ref{LA})-(\ref{SP}). The results
  for $b_n(Q_0^2)$ from (\ref{BMS}), (\ref{MPS}), and (\ref{SP}) are given in brackets, wavy brackets, and square brackets, respectively.
}
\label{Tab:Ind}
\end{table}

\begin{table}[t]
\begin{center}
\begin{tabular}{|c|c|c|c|}
\hline
& $M^2$ [GeV$^2$] & $\mu_{\rm{A},4}(Q^2_0)$  & $\chi^2/({\rm d.o.f.})$   \\
\hline 
LO & 1.647 $\pm$ 0.184 (1.647 $\pm$ 0.184) & -0.161 $\pm$ 0.006 (-0.161 $\pm$ 0.006) & 0.41 (0.41)  \\
& \{1.647 $\pm$ 0.184\} [1.647 $\pm$ 0.184] & \{-0.161 $\pm$ 0.006\} [-0.161 $\pm$ 0.006] & \{0.41\} [0.41]  \\
\hline
  NLO & 0.565 $\pm$ 0.123 (0.567 $\pm$ 0.124) & -0.162 $\pm$ 0.019 (-0.162 $\pm$ 0.019) & 0.52 (0.52) \\
 & \{0.564 $\pm$ 0.123\} [0.562 $\pm$ 0.123] & \{-0.162 $\pm$ 0.019\} [-0.162 $\pm$ 0.019] & \{0.52\} [0.52] \\
  \hline
  N$^2$LO & 0.209 $\pm$ 0.095 (0.211 $\pm$ 0.095) & -0.221 $\pm$ 0.073 (-0.220 $\pm$ 0.073)& 1.12 (1.12)  \\
  & \{0.209 $\pm$ 0.095\} [0.207 $\pm$ 0.095] & \{-0.221 $\pm$ 0.073\} [-0.221 $\pm$ 0.074])& \{1.13\} [1.13]  \\
 \hline
\end{tabular}
\end{center}
\vspace{-0.3cm}
\caption{%
The same as in Table \ref{Tab:Ind}, but for $Q^2$-dependent $\mu_{\rm{A},4}$ from Eq. (\ref{Tw4A}).
}
\label{Tab:Dep}
\end{table}

In the case of APT, we observe good agreement between the QCD predictions and experimental data (see Fig. 2 and Tables 1 and 2).
Looking at Tables 1 and 2, we see that the obtained results are quite similar and independent of the choice of $b_2$ and $b_4$. This may apparently be due to the
fact that the values of $b_2$ and $b_4$ are extracted in the combination $b_2+b_4$ (see \cite{Bakulev:2003cs} and the discussion therein).
Indeed, the results for $b_2+b_4$ are very close to each other in the cases (\ref{LA}), (\ref{BMS}), and (\ref{SP}).
For the set MPS$_2$, this sum is larger, but the results in Tables 1 and 2 are very similar.
This requires further research.

In the case where the massive twist-four term can be expanded in an inverse series in $Q^2$, it can be represented as
\be
\frac{k_{\rm A,4}(Q^2)}{Q^2}+\frac{k_{\rm A,6}(Q^2)}{Q^4}~~(k_{\rm A,4}=\mu_{\rm A,4}(Q^2)M^{2},~~k_{\rm A,6}=-\mu_{\rm A,4}(Q^2)M^{4})\,.
\label{kn}
\ee
The values of $k_{\rm A,4}(Q_0^2)$ and especially $k_{\rm A,6}(Q_0^2)$ decrease strongly with increasing PT order
(see the same conclusions on other topics in Refs. \cite{Kataev:1996vu,Kataev:1999bp}), which also demonstrates the good applicability of APT.
Moreover, the next-to-leading (NLO) result for $k_{\rm A,4}(Q_0^2)$
\be
k^{\rm NLO}_{4}(Q_0^2)=-0.92\pm 0.023~ \mbox{GeV}^2\,,
\label{k4}
\ee
given in Table 2 is in complete agreement with the NLO estimate 
\cite{Bakulev:2002uc}
\be
k_{4}(Q_0^2)=-\frac{80\sqrt{2}f_{\pi}}{27}\delta^2(Q_0^2)=-0.104\pm 0.011~ \mbox{GeV}^2~~(\delta^2(Q_0^2)=0.19\pm 0.02~ \mbox{GeV}^2)\,.
\label{k4MPS}
\ee

\section{Conclusions}

We have investigated the pion-photon TFF
within the framework of conventional and analytic QCD.
A comparison was performed between experimental data
and perturbative QCD
based on PT and APT
with the ``massive'' form of
the twist-four contribution. We showed that conventional PT 
beyond LO
fails to reproduce the data, while
the analytic version
demonstrates good agreement with experimental data. 

  The results obtained in APT
  are very stable with increasing PT order. Moreover, they are in perfect agreement with the results in \cite{Mikhailov:2021znq} (see Fig. 2),
  obtained in the framework of so-called method of lightcone sum rules in forms of disperssion relations \cite{Balitsky:1989ry} (see also \cite{Ayala:2018ifo}).

As can be seen from the discussions in Refs. \cite{Mikhailov:2021znq,Ayala:2018ifo}, the most appropriate choice of scales is $\mu_f\sim \mu_r \sim Q^2x$.
We plan to consider this choice of scales in our future research.\\


{\bf Acknowledgments.}~
One of us (I.A.Z.) was supported in part by the Fellowship ANID
Beca de Doctorado Nacional No. 21250067. A.V.K.
was supported  by  the Russian Science Foundation grant No. 25-22-00576.

\end{document}